\documentclass{article}

\usepackage{PRIMEarxiv}

\usepackage[utf8]{inputenc} 
\usepackage[T1]{fontenc}    
\usepackage{hyperref}       
\usepackage{url}            
\usepackage{booktabs}       
\usepackage{amsfonts}       
\usepackage{nicefrac}       
\usepackage{microtype}      
\usepackage{fancyhdr}       
\usepackage{graphicx}       
\usepackage{amsmath}
\usepackage{amssymb}
\usepackage{float}
\usepackage{subcaption}
\usepackage{orcidlink}

\title{Belief-Based Maximum Occupancy Principle and\\ Active Inference}

\author{
  Manolis Mylonas\,\orcidlink{0009-0001-8890-0630} \\
  Center for Brain and Cognition \& Department of Engineering \\
  Universitat Pompeu Fabra \\
  Barcelona, Spain \\
  \texttt{manolis.mylonas@upf.edu} \\
   \And
  Rub\'{e}n Moreno-Bote\,\orcidlink{0000-0003-0398-014X} \\
  Center for Brain and Cognition \& Department of Engineering \\
  Universitat Pompeu Fabra \\
  Barcelona, Spain \\
  \texttt{ruben.moreno@upf.edu} \\
}

\begin{document}
\maketitle

\newcommand\blfootnote[1]{%
  \begingroup
  \renewcommand\thefootnote{}\footnote{#1}%
  \addtocounter{footnote}{-1}%
  \endgroup
}

\blfootnote{This version of the contribution has been accepted for
publication at the 7th International Workshop on Active Inference
(IWAI 2026), after peer review, but is not the Version of Record and
does not reflect post-acceptance improvements, or any corrections.
The Version of Record will be published in the IWAI 2026 proceedings
(Springer, Communications in Computer and Information Science).
Use of this Accepted Version is subject to the publisher's Accepted
Manuscript terms of use:
\url{https://www.springernature.com/gp/open-research/policies/accepted-manuscript-terms}}

\begin{abstract}
Intrinsic motivation plays a central role in adaptive and goal-directed behavior by conferring agents reward-independent objectives and biases useful to act in noisy and uncertain environments.
Active Inference addresses the problem of acting in a partially observable environment through a principled framework for belief updating and action selection. A key component of Active Inference is the specification of prior preferences, which shapes behavior by encoding desirable future outcomes. 
An intrinsic motivation approach called the Maximum Occupancy Principle (MOP) proposes that agents act so as to maximize occupancy over future paths of states and actions, with no preferences or epistemic targets. Despite its simple formulation, MOP gives rise to rich and adaptive behaviors that combine exploratory variability with goal-directed dynamics. 
In this work, we extend MOP to partially observable environments and introduce a Bellman reformulation of the Expected Free Energy for Active Inference, both incorporating belief-based inference over hidden states as part of the agent state. The Bellman formulation enables tractable offline computation via value iteration over the full belief-state space. We compare the resulting behaviors in a set of minimal experimental settings with uncertain food sources. 
We find that MOP agents switch between goal-directed (food seeking) behavior and exploration between different food sources, depending on their energy available and their belief state. In contrast, Active Inference agents mostly inhabit regions around a single food source, a strategy having both high pragmatic and epistemic value. We finally compare with Empowerment, which is shown to be qualitatively similar to Active Inference.  

\end{abstract}

\section{Introduction}
While extrinsic reward maximization --the idea that there is a fixed reward function that must be maximized by the agent in the long run-- has dominated reinforcement learning \cite{sutton1998reinforcement}, many behaviors observed in biological agents cannot be fully explained through reward optimization alone. Curiosity-driven exploration, spontaneous action selection, and behavioral variability are often attributed to intrinsic motivational mechanisms that operate independently of extrinsic rewards \cite{moreno2026intrinsic}. Active Inference \cite{smith2022step,da2023reward,friston2021sophisticated} is a prominent theory of decision making not purely based on extrinsic rewards. In this framework, the agent aims to reach states that provide information about the environment, but also agree with some prior preferences that ensure homeostasis, by minimizing its free energy.  
Another popular theory is Empowerment \cite{jung2011empowerment,salge2014empowerment}, which offers an alternative information-theoretic perspective, quantifying the influence an agent has on its environment by maximizing the mutual information between actions and future states. This yields behavior that favors controllability and observability of the environment by the agent. A more recent intrinsic motivation theory, the Maximum Occupancy Principle (MOP) \cite{ramirez2024complex,moreno2023empowerment}, departs from reward-based formulations by proposing that agents should act to maximize the entropy of future action-state paths: actions and states are preferred when they support a diverse set of possible future trajectories, thereby maintaining behavioral flexibility and adaptability. The resulting control problem can be expressed through a value function that quantifies the expected diversity of future paths under a given policy. While simple in formulation, this approach can generate complex behaviors, combining both variability and goal-directedness without relying on explicit rewards; it also produces life-preserving behaviors to avoid falling in terminal states from where no more path entropy can be generated \cite{moreno2023empowerment}. 

Partially observable Markov decision processes (POMDPs) provide the standard formalism for decision-making under uncertainty \cite{smallwood1973optimal,sondik1971optimal,zhang2010partially,lovejoy1991computationally,lovejoy1991survey}. In this setting, agents do not directly observe the full system state but instead receive noisy observations, requiring the construction of a belief over hidden variables. This belief evolves in time as a stochastic process and acts as a sufficient statistic for optimal control \cite{sondik1978optimal}.
MOP was originally formulated for fully observable environments. However, most realistic settings involve partial observability. In this work, we extend MOP to POMDPs by introducing belief-based inference over hidden variables and deriving the corresponding value function (functional) and optimal policy. We demonstrate the resulting behavior in a two-food-source partially observable grid-world environment and compare it with Active Inference and Empowerment. Our results show that MOP induces adaptive exploration strategies between and around the food sources that are strongly modulated by internal energetic constraints. 
We also introduce two extensions of the Sophisticated Inference framework: we reformulate the Expected Free Energy as a Bellman equation with a temporal discount factor, replacing the original tree-search procedure with tractable offline value iteration over the full belief-state space, and we derive time-stationary policies that do not require replanning at each decision step.
Both Active Inference and Empowerment agents tend to inhabit a single food source, where higher epistemic value can be gained and also larger controllability can be exerted. 
While the three frameworks combine exploratory and homeostatic pressures in their induced behavior, they do so in fundamentally different ways. This comparison is part of a broader effort to build a theory of behavior: we compare the three frameworks in terms of the richness and spatial extent of the behavior they induce, and in terms of whether goal-directed action emerges without being explicitly encoded.

\section{Partial Observability Formulation}

\begin{figure}[t]
\centering
\includegraphics[width=0.6\textwidth]{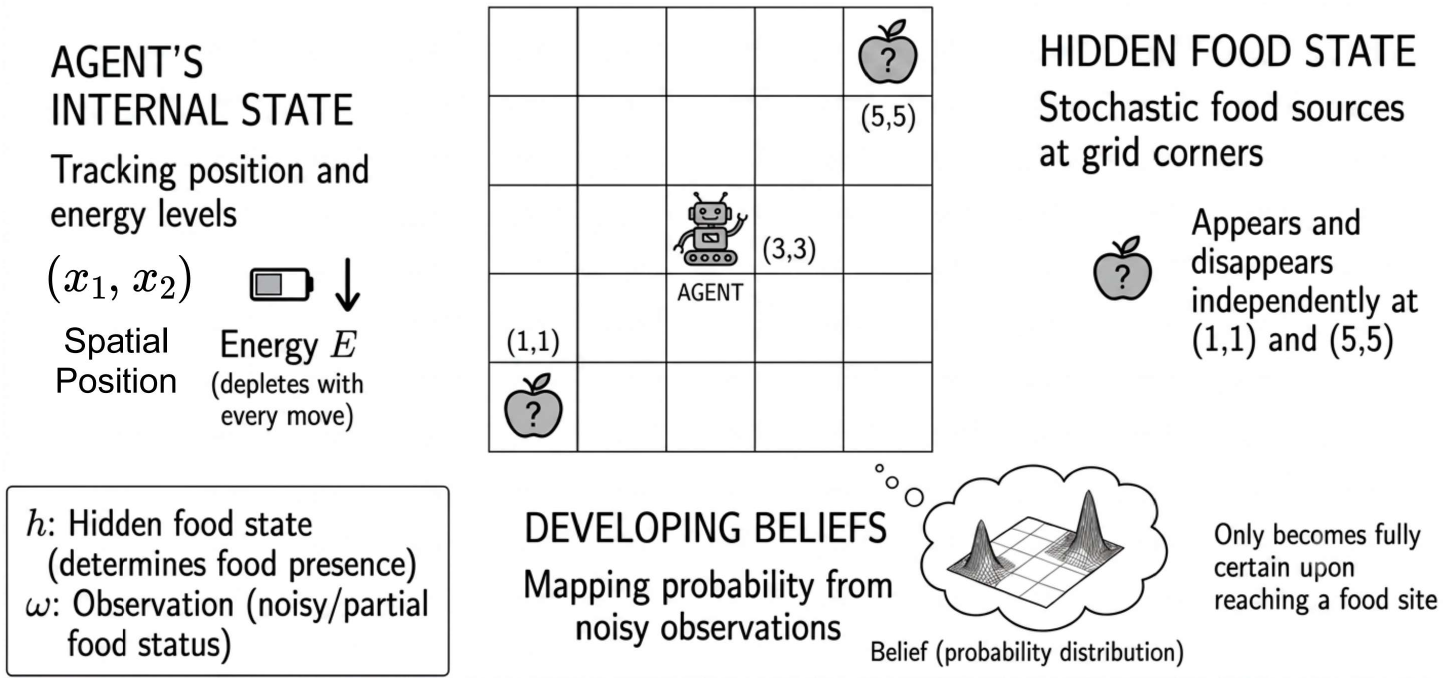}
\caption{Two-Food-Source Partially Observable Grid-World Environment}
\label{fig1}
\end{figure}

Although our framework is general and applies to any partially observable Markov decision process (POMDP), here we specialize the notation and show results for the particular case of a two-food-source partially observable grid-world (Fig. \ref{fig1}). Extending our equations and notation to any general POMDP is straightforward, and the exact general update equations for MOP, EFE and Empowerment are given in Eqs. \ref{v_pomdp_opt}, \ref{EFE} and \ref{mpow}, respectively.
Our specific environment includes the following key ingredients, enabling a systematic comparison between different intrinsic motivation approaches: (1) multiple sources of evidence, allowing us to assess whether different methods preferentially focus on a single source or distribute attention across both; (2) an internal energy variable, which permits evaluation of how conservatively each approach behaves with respect to avoiding low-energy (terminal) states; and (3) a spatial separation between food sources, which makes it possible to measure the extent of exploratory excursions and the ability of each method to sustain long-range exploration.

\subsection{A two-food-source partially observable grid-world environment}

We consider a POMDP in discrete time defined by the tuple
$M = \{S,\allowbreak\, H,\allowbreak\, E_{\mathrm{gain}},\allowbreak\, E_{\max},\allowbreak\, A,\allowbreak\, \Omega,\allowbreak\, p,\allowbreak\, q\}$.
The fully observable state space is 
$S=\mathcal X_1 \times \mathcal X_2 \times \mathcal E$,
where $x = (x_1,x_2)\in\mathcal X_1\times\mathcal X_2$ denotes the agent's position in a $5\times5$ grid world and $E\in\mathcal E$ its energy level. The environment contains two food sources $i \in \{ 0,1\}$, located at positions $x_{f,i=1}=(1,1)$ and $x_{f,i=2}=(5,5)$.
Their availability is represented by the hidden state
$H=\{h_1,h_2\}$,
where $h_i\in\{0,1\}$ is a binary variable indicating whether food source $ x_{f,i}$ contains food ($h_i = 1$) or not ($h_i = 0$). We denote the observable state by
\[
s=(x,E)=(x_1,x_2,E)
\]
and the hidden state by
\[
h=(h_1,h_2) \;.
\]
The action space is
$A=\{\mathrm{up},\mathrm{down},\mathrm{left},\mathrm{right},\mathrm{stay}\}$. A table with all the definitions of the terms used throughout and the implementation values can be found in Appendix \ref{app:definitions} and \ref{app:implementation} respectively.

\setlength{\textfloatsep}{8pt plus 1pt minus 1pt}
\subsection{Environment Dynamics} \label{dynamics}

The probability transition matrix of the world (observable and hidden) state $(s,h)$ to its successor $(s',h')$, given that action $a$ is performed, is
\begin{equation}\label{env}
p(s',h'|s,h,a)
= p(E'|E, x', h') \; p(x'|s,a) \; p(h'|h, x) \;.
\end{equation}
We assume that the transition matrix factorizes into
\begin{equation*}
p(s',h'|s,h,a)=
\underbrace{\delta\bigl(E'-E'(E, x',h')\bigr)}_{\text{energy transition}}
\;\underbrace{\delta \bigl(x'-x'(s,a)\bigr)}_{\text{spatial transition}}
\; \underbrace{p(h_1'|h_1,x) \; p(h_2'|h_2,x)}_{\text{food transition}} \;,
\end{equation*}
which models a scenario where food sources evolve independently of each other, and whose state might depend on the visitation of the agent to the food source location, but not on the agent's energy. 
Here, $\delta$ is the Kronecker delta, and we write $x'(s,a)$ and $E'(E, x', h')$ for the resulting successor position and energy, $s'=(x',E')$. The spatial transition is deterministic given the current state and action. 
The agent moves one cell in the chosen direction, subject to the boundary constraints of the grid-world and the availability of energy,
\begin{equation*}
x'(s,a) \equiv x'(x,E,a) =
\begin{cases}
x + \Delta(a) & \text{if } x + \Delta(a) \text{ lies within the grid and } E > 0,\\
x & \text{otherwise,}
\end{cases}
\end{equation*}
where $\Delta(a)$ is the displacement associated with action $a$. In 
particular, if the agent reaches zero energy, $E = 0$, it "dies" and cannot 
move, i.e., the only available action is to "stay" with probability one. Thus, the resulting state has zero action entropy. The energy transition depends not only on the current energy but also on the new position $x'$ and the new food state $h'$ deterministically as
\begin{equation*}
E'(E, x',h') =
\begin{cases}
\min \;\!\bigl(E - 1 + E_{\mathrm{gain}},\; E_{\max}\bigr) & \text{if } x' =x_{f,i} \text{ and } h_i' = 1,\\
E - 1 & \text{otherwise,}
\end{cases}
\end{equation*}
since the agent can only gain energy if it lands on a food source where food is present.
This makes explicit that the spatial dynamics are autonomous (a function of $s$ and $a$ alone), whereas the energy dynamics are coupled to the hidden food state through $h'$. 

For each food source, we consider two possible dynamics:
\paragraph{Location-independent dynamics.} Here, $p(h_i'|h_i,x) = p(h_i'|h_i)$, independent of the agent's location, with
\begin{align*}
p(h_i'=0|h_i=1)&=1-\lambda \;, &
p(h_i'=0|h_i=0)&=1-\mu \;.
\end{align*}
\paragraph{Depletion at visit.}
Here, $p(h_i'|h_i,x)$ explicitly depends on the agent's location,
\begin{gather*}
p(h_i'=0|h_i=1,x\neq x_{f,i}) = 1-\lambda, \qquad p(h_i'=0|h_i=1,x=x_{f,i}) = 1-\rho,\\
p(h_i'=0|h_i=0,x) = 1-\mu \;, \; \; \forall  x \;,
\end{gather*}
where $\mu, \lambda, \rho \in [0,1]$ are parameters that determine the stochasticity of the food-state transitions. 

\subsection{Observation Model}

The agent receives noisy binary observations $\omega=(\omega_1,\omega_2)$ related to the presence or absence of food at each food source location. The observation model is
\begin{equation}\label{obs}
q(\omega|s,h)=
q(\omega_1|x,h_1)\,
q(\omega_2|x,h_2) \;,
\end{equation}
assuming conditional independence of the observations given the agent's location $x$ and hidden state $h$, and no dependence on the agent's current energy.
We assume that the observation is fully informative at the food location, and totally uninformative away from it, that is
\begin{equation*}
q(\omega_i=h_i | x=x_{f,i}, h_i)=1, \qquad q(\omega_i= \{1,0\} | x\neq x_{f,i}, h_i)=\tfrac{1}{2} \;.
\end{equation*}
This means e.g. that observation $\omega_i=1$ at the food source implies with certainty that the food source $i$ is full, $h_i=1$. In contrast, outside the food source, observations are noisy and non-informative. This choice is made to keep a simple representation of the environment while preserving the structure of partial observability.

\subsection{Belief-State Representation} \label{section_beliefs}
We define the belief of the agent about the hidden variable $h$, and denote it as $b(h)$, as the probability distribution over $h$ inferred from all past observations, $b(h) \equiv P(h|\text{past observations})$ 
(we often denote this probability distribution simply by $b$). 
Starting from belief $b$, an agent at observable state $s=(x,E)$ performs an action $a$, experiences a transition to new location $x'$ and observes $\omega'$. After this observation, the agent can update the belief about the hidden variables using Bayes rule as 
\begin{equation}\label{belief}
b'(h') \equiv P(h'|\omega',s,b,a)
\propto
\sum_{x',h} q(\omega'|x',h')\; p(h'|h, x)\;\delta(x' -x'(s,a)) \;b(h) \;.
\end{equation}
We will use further below the notation $b'(\omega',s,b,a)$ to indicate this updated belief (probability distribution) and make explicit all its dependencies.
Because both the observation and the hidden state dynamics are conditionally independent given the world state $(s,h)$ (see Eqs. \ref{env},\ref{obs}), if the initial belief is factorized, then the updated belief is also factorized (note that initially the agent can start with a belief that is factorized to represent the real world state of food source independence, which is what we assume here). Thus, the belief factors in $b'(h') = b'(h_1') b'(h_2')$ can be computed as
\begin{equation*}
b'(h'_i) \equiv P(h_i'|\omega'_i,s,b,a)
\propto
\sum_{h_i} q(\omega'_i|x'(s,a),h_i')\;p(h_i'|h_i,x)\;b(h_i)
.
\end{equation*}
Note that belief factorization holds in both the location-independent and the depletion-at-visit environments. 

In general, the belief $b$ is a continuous quantity, as it represents a probability distribution over the hidden state $h \in \{0,1\}$ and can take any value in $[0,1] \times [0,1]$. For computational tractability, however, we discretize the belief state, so that the full agent state $(s,b)$ lies on a finite grid and standard dynamic programming methods can be applied (see Appendix \ref{app:beliefs} and \ref{app:beliefs_simple} for details).

\subsection{Predictive Model}
The agent uses a predictive model during planning by first replacing the unknown hidden states by beliefs. The sufficient statistics for the problem is the agent state $(s,b)$, which summarizes all information available for the agent to compute an optimal policy and select actions. 
Optimal policies depend on the optimization objective, which will be described in detail in Secs. \ref{MOP}-\ref{FEP}-\ref{EMP} for MOP, EFE and Empowerment, respectively.

The probability distribution of the next agent's state $(s',b')$ given $(s,b)$ and the performed action $a$ is computed as
\begin{equation} \label{predictive_sb}
    P(s',b' |s,b,a) =  \sum_{\omega'} \delta(b'-b'(\omega',s,a)) P(s',\omega'|s,b,a) \;,
\end{equation}
with $P(s',\omega'|s,b,a) =  \sum_{h,h'} q(\omega'|x'(s,a),h') p(s',h'|s,h,a) b(h)$, where we have used the updated belief in Eq. \ref{belief}, the observation model in Eq. \ref{obs}, and the world (observable and hidden) state transition dynamics in Eq. \ref{env}. 
Note that the belief evolves deterministically, a fact that is represented by the Kronecker delta function (it is one if the probability distribution $b'$ matches $b'(\omega',x,a)$, and zero otherwise, noting again that probabilities are discretized).
 Eq. \ref{predictive_sb} is used in the definitions of MOP and Active Inference in Eqs. \ref{v_pomdp_opt} and \ref{EFE_action}. 

The full predictive model is the joint probability distribution over observations, future observable states and hidden states, given the current agent's state $(s,b)$ and performed action $a$. This joint, used in Eq. \ref{EFE_action}, is expressed as
\begin{equation} \label{predictive_hidden}
    P(\omega', s', h'| s, b, a) =  \sum_h q(\omega'|x',h') p(s',h'|s, h, a) b(h) \;.
\end{equation}

\section{Intrinsic motivation models}
\subsection{Maximum Occupancy Principle} \label{MOP}

In POMDPs, the policy can only depend on the observable state $s$ and the current belief of the agent $b$ about the hidden states, and not on the hidden states $h$ themselves. 
Indeed, and as we said before, the agent state $(s,b)$ corresponds to the sufficient statistics in our problem. 
We define the policy $\pi(a|s,b)$ as the probability of selecting action $a$ given the observable state $s$ and belief $b$. Starting at $t=0$ in state $(s_0,b_0)$, an agent performing a sequence of actions and experiencing state transitions $\tau = (s_0, b_0, a_0, s_1,b_1, ..., a_t,s_{t+1},b_{t+1},...)$ gets the return (intrinsic reward)
\begin{equation*}
R(\tau)
=
\sum_{t=0}^{\infty}
\gamma^{t} R(s_t,b_t,a_t)
=
-
\sum_{t=0}^{\infty}
\gamma^t
\ln
\left(
\pi(a_t| s_t,b_t)
\,
P^{\beta}(s_{t+1},b_{t+1}| s_t,b_t,a_t) 
\right) \; ,
\end{equation*}
with $\beta$ being a fixed real number (chosen for simplicity to be $\beta=0$ in our simulations), and discount factor $0<\gamma<1$. The corresponding value function (functional) in the POMDP setting is given by
\begin{equation*}
V_{\pi}(s,b)
=
\mathbb{E}_{\pi,P}
\left[
\sum_{t=0}^{\infty}
\gamma^t
\left(
H(a|s_t,b_t)
+
\beta H(s',b'|s_t,b_t,a_t)
\right)
\;\middle|\;
s_0=s, b_0=b
\right] \;,
\end{equation*}
where $H(a|s_t,b_t)$ is the policy entropy, and $H(\cdot,\cdot|s_t,b_t,a_t)$ is the entropy of the agent state transition kernel $P(s',b'|s,b,a)$ defined in Eq. \ref{predictive_sb}.
The optimal policy is such that the value function is optimized in every agent's state. 
This optimal value function $V^*(s,b)$ satisfies the self-consistency equation
\begin{equation}
\label{v_pomdp_opt}
V^*(s,b)
=
\ln
\sum_a
\exp
\left[
\beta H(\cdot,\cdot|s,b,a)
+
\gamma
\sum_{s', b'}
P(s', b'|s,b,a)
V^*(s',b')
\right] \;.
\end{equation}
The sum over $b'$ is taken to be a finite sum rather than an integral, because we discretize belief state onto a finite grid (Sec. \ref{section_beliefs}).
This equation can be solved by iterative recursion \cite{ramirez2024complex}, where we have to impose the boundary conditions $V(s^+,b)=0$ for every observable state $s^+$ with $E=0$ regardless of $b$ so that terminal states have zero value due to the termination of the episode and the impossibility of generating any further action-state path entropy. Implementation details can be found in Appendix \ref{implementation}. 

The optimal action-value function is defined as
\begin{equation} \label{optimal_q}
Q^*(s,b,a) = \gamma \sum_{s', b'} P(s', b'|s,b,a)\, V^*(s',b') \;,
\end{equation}
from where the optimal policy is computed as
\begin{equation} \label{optimal_mop_policy}
\pi^*(a|s,b) = \sigma\big(Q^*(s,b,a)\big) = \frac{\exp\left[ Q^*(s,b,a)\right]}{\sum_{a'}\exp\left[ Q^*(s,b,a')\right]} \;,
\end{equation}
where $\sigma$ denotes the softmax operation.

\subsection{Sophisticated Inference} \label{FEP}
The implementation of Active Inference builds on Sophisticated Inference \cite{friston2021sophisticated}, with the novelty of including exact belief propagation and temporal discounting, which enables time-stationary solutions. In the Active Inference framework, the prior preferences over the location-energy state $s$ are embodied in a target distribution $P(s)$, which is taken here to be uniform across states except for $E=0$, which is assigned a low probability (low preference). 
With this choice, we aim at introducing the least structure into the agent's preferences. 

\subsubsection{Variational Free Energy}
The agent forms beliefs by minimizing its Variational Free Energy (VFE). The VFE over hidden food states $h$ is defined as
\begin{equation}\label{VFE}
    F[Q] = D_{\mathrm{KL}}\big(Q(h')\,\|\,p(h')\big)
    - \mathbb{E}_{Q(h')}\big[\ln q(\omega' | s', h')\big] \;,
\end{equation}
where $Q(h')$ is the variational distribution over the updated hidden food state $h'$ obtained after taking action $a$ and transitioning from state $s$ to $s'$, $p(h') = \sum_h p(h'|h, x) \; Q(h)$ is the predictive prior over $h'$ obtained by propagating the current variational belief $Q(h)$ through the transition dynamics, and $q(\omega' | s', h')$ is the observation model defined in Eq.~\ref{obs}. The first term penalizes deviations from the prior (complexity), while the second term encodes the fit to observations (accuracy). Minimizing the VFE in the grid-world scenario is mathematically equivalent to the belief propagation equations described in Sec.~\ref{section_beliefs} (see Appendix~\ref{VFE_belief} for details) and the belief $b(h)$ used throughout this work corresponds to the minimizer $b(h) = \arg\min_Q F[Q]$. In the following section, the beliefs used in the computation of the Expected Free Energy are obtained through the same belief update mechanism employed in MOP (Eq.~\ref{belief}).

\subsubsection{Expected Free Energy}
In the Sophisticated Inference framework, the agent selects actions so as to minimize its Expected Free Energy (EFE), which balances pragmatic drives (risk) with epistemic drives (ambiguity resolution). The EFE is written as
\begin{align} \label{EFE}
    G(s,b)
&=
\sum_a \pi(a|s,b) \;
G(s, b, a) \;,
\end{align}
where $G(s,b,a)$ is the per-action EFE, defined as
\begin{align}\label{EFE_action}
    G(s,b,a)
&=
\underbrace{
\mathbb{E}_{P(\omega', s', h'|s, b, a)}
\Big[
\overbrace{
\ln P(s'|s, b, a)
-\ln P(s')
}^{\text{Risk}}
\;
\overbrace{
-\ln q(\omega'|s', h')
}^{\text{Ambiguity}}
\Big]
}_{\text{Expected free energy of next action}}
\notag
\\
&\quad
+\gamma
\underbrace{
\mathbb{E}_{P(s', b'|s, b, a)}
\big[
G(s', b')
\big] \;.
}_{\text{Expected free energy of subsequent actions}}
\end{align}
The policy $\pi(a|s, b)$ is computed as
\begin{align} \label{G_policy}
    \pi(a|s,b) = \sigma\big(-d \; G(s,b,a)\big) \;,
\end{align}
where $d$ is the inverse temperature parameter. The negative sign ensures that actions with lower EFE are assigned higher probability, consistent with the agent's objective of minimizing $G(s,b,a)$. Since beliefs $b$ are discretized onto a finite grid (Sec. \ref{section_beliefs}), expectations become finite sums, making $G(s,b)$ tractable to compute via value iteration over the full state-belief space $(s,b)$.
The choice of terminal boundary condition $G^+$ is important, since it governs the trade-off between exploration and conservatism in an active inference agent; this is discussed in detail in Appendix~\ref{app:bound_cond}.

We can also recover the optimal policy by taking the limit $d \to \infty$ in Eq. \ref{G_policy}, which reduces the softmax to a hard minimization over actions (and thus a deterministic policy, except for ties), replacing Eq.~\ref{EFE} with
\begin{equation} \label{optimal_EFE}
    G(s,b)
=\min_a
G(s, b, a) \;.
\end{equation}

In the deterministic (optimal) EFE (Eq. \ref{optimal_EFE}), the probability mass concentrates entirely on the action that minimizes $G(s,b,a)$. In the stochastic EFE (Eqs. \ref{EFE}-\ref{G_policy}), $d$ is a tunable parameter controlling the degree of randomness in the policy. The effect of $d$ on the agent's behavior is analyzed in Results and Appendix ~\ref{app:temperature}. 

Eqs.~\ref{EFE}-\ref{G_policy} have the form of a Bellman equation and can be solved via value iteration \cite{sutton1998reinforcement}. In contrast to \cite{friston2021sophisticated}, where the EFE is computed by a deep tree search whose cost scales exponentially with depth, this allows the EFE to be computed offline over the full belief-state space with cost linear in depth, and the discount factor $\gamma$ guarantees a unique fixed point and a time-stationary policy. An interpretation of the risk and ambiguity terms can be found in Appendix \ref{app:EFE_explain}.

\subsection{Empowerment} \label{EMP}

Empowerment is an intrinsic objective that encourages the agent to select actions that maximize its influence over future observations. Specifically, it measures the mutual information between an $n$-step action sequence and the resulting future observation, quantifying how much control the agent has over what it will perceive. Implementation details can be found in Appendix ~\ref{app:mpow}.

\section{Results}


\setlength{\textfloatsep}{8pt plus 1pt minus 1pt}
\begin{figure}[t]
\centering
{\small\textbf{Independent food transition dynamics}}\\[0.3em]
\includegraphics[width=0.7\linewidth]{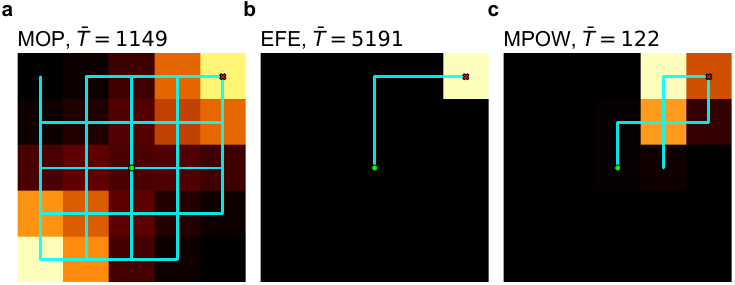}
\vspace{0.1cm}

{\small\textbf{Food depletion transition dynamics}}\\[0.3em]
\includegraphics[width=0.7\linewidth]{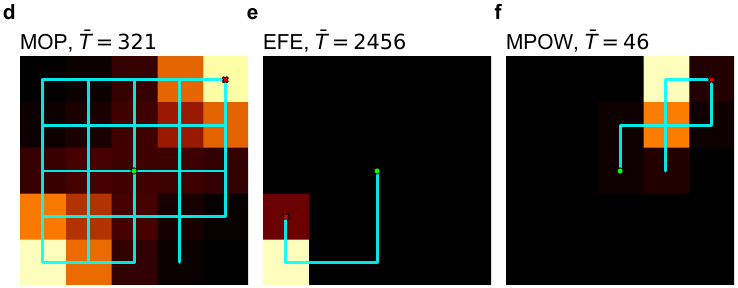}
\caption{Space habitation heatmaps (average over 100 episodes) and a sample trajectory for MOP, EFE 
and Empowerment (MPOW), under independent (top) and 
depletion (bottom) food dynamics. $\bar{T}$ denotes average survival 
duration in steps.}
\label{fig:comparison}
\end{figure}

We first evaluate and compare MOP (Eq. ~\ref{v_pomdp_opt}), Active Inference (Eqs. \ref{EFE}-\ref{G_policy}) and Empowerment (Eq. ~\ref{mpow}). For each method, we compute the corresponding objective,
we derive the induced policy and simulate trajectories from identical initial conditions (Fig.~\ref{fig:comparison}).
The resulting behaviors exhibit distinct qualitative structures: \\
\textbf{$\cdot$ MOP}. The induced policy explores the environment extensively, visiting both food sources and frequently switching between them. The resulting paths show sustained exploration without convergence to a single attractor. \\
\textbf{$\cdot$ Active Inference}. The policy converges to a single food source and remains in its vicinity. This behavior is consistent with the minimization of expected free energy, which jointly favors energy preservation and reliable observations, leading to a stable attractor state. The agent records high average survival steps, at the expense of exploration. Here the deterministic policy of EFE is used (using a high inverse temperature $d$ in Eq. ~\ref{G_policy}). A comparison between stochastic and deterministic EFE is provided in Appendix ~\ref{app:temperature}\\
\textbf{$\cdot$ Empowerment}. The policy reaches a food source and performs limited local exploration. Due to the finite planning horizon ($n=5$), the agent does not systematically evaluate distant alternatives, resulting in localized rather than global exploration behavior. A higher planning horizon (e.g., $n=10$) that would allow the agent to see the other food state is significantly more computationally expensive than the other two methods, and thus is not considered. The policy of empowerment is by definition deterministic, as the one maximizing the mutual information.

\begin{figure}[!t]
\centering
\includegraphics[width=0.7\columnwidth]{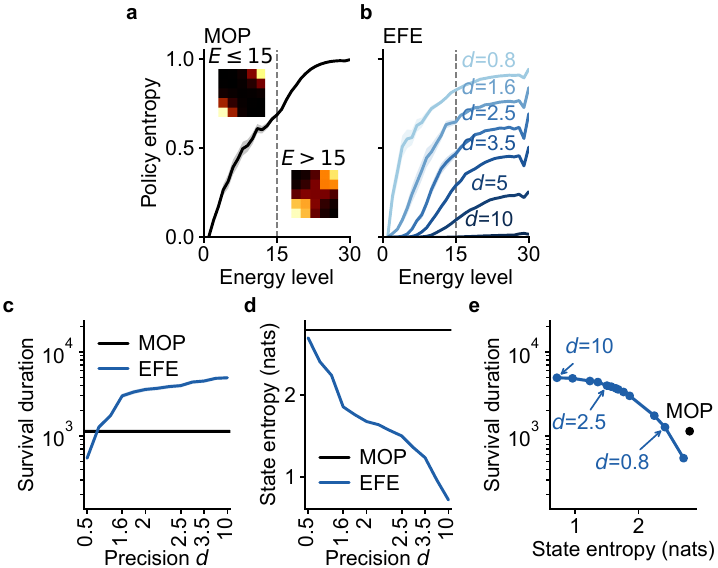}
\caption{Normalized policy entropy, survival and exploration for MOP and Active Inference, in the location-independent dynamics; depletion-at-visit shows similar behavior.
 \textbf{(a)} MOP exhibits a transition when the agent's internal energy is around $E=15$, switching between broad exploration at high energy and goal-directed behavior at low energy (insets are corresponding state visitation heatmaps). \textbf{(b)} Active Inference policy entropy for six values of the inverse temperature $d$. Every curve increases with energy, but $d$ modulates the behavior: for small $d$ the policy entropy remains high and the policy stays stochastic, whereas for large $d$ it falls toward zero at every energy level and the policy becomes nearly deterministic (see heatmaps in Appendix~\ref{app:temperature}). \textbf{(c)} Average survival duration of the Active Inference agent as a function of $d$, compared with MOP. \textbf{(d)} Entropy of the empirical state-visitation distribution over the grid, over the same range of $d$, compared with MOP. Panels (c) and (d) are approximately mirror images of each other: the values of $d$ at which its occupancy of the environment
collapses \textbf{(e)} The same trade-off as a single curve, survival duration against the entropy of the empirical state-visitation distribution over the grid, one point per value of $d$, with MOP as a single point lying off the trace, indicated best survival for matched fixed state entropy, or best state entropy for matched fixed survival.}
\label{fig:entropy_energy}
\end{figure}

We next analyze how MOP modulates behavioral variability as a function of the agent's energy $E$. To this end, we compute the (normalized) entropy of the induced policy for different energy levels $E$ and examine the corresponding state visitation patterns (Fig.~\ref{fig:entropy_energy}a). The value $E=15$ roughly separates two energy-dependent regimes in the policy structure. For high energy levels ($E>15$), the agent exhibits increased policy entropy, corresponding to a broader distribution over actions. In this regime, the agent can afford exploratory behavior, leading to more uniform state visitation and frequent switching between food sources. In contrast, for low energy levels ($E\le 15$), the policy becomes significantly more concentrated, similar to the deterministic EFE and Empowerment agents. MOP prioritizes reaching food sources to avoid starvation, resulting in reduced entropy and more directed trajectories toward high-reward states. Overall, these results indicate that MOP induces an adaptive control strategy in which internal energetic constraints modulate the level of stochasticity in the policy. 

Active Inference exhibits the same qualitative pattern (Fig.~\ref{fig:entropy_energy}b). The policy entropy is highest when the agent is well fed and decreases as its energy depletes, so that behavioral variability is again modulated by the internal energetic state. What differs is the level at which this modulation operates, and that level is set by $d$ rather than by the agent: the larger the inverse temperature, the lower the entropy at every energy level. Because $d$ is fixed externally rather than modulated by the agent's own energetic state, it acts as a single control knob that trades exploration against survival (Figs.~\ref{fig:entropy_energy}c-e). For small $d$ the agent is adventurous: it spreads its visits over most of the grid, reaching a state entropy close to MOP's (Fig.~\ref{fig:entropy_energy}d), but it attains the shortest lifespans of the range (Fig.~\ref{fig:entropy_energy}c). As $d$ increases the agent becomes conservative: it settles on a single food source, the location that simultaneously minimizes risk and ambiguity in Eq.~\ref{EFE_action}, and remains in its vicinity. State entropy falls from $2.7$ to $0.7$ nats between $d=0.5$ and $d=10$, while survival rises monotonically with $d$. Survival is therefore bought by abandoning exploration, and the exchange rate is set by $d$ alone. A high policy entropy does not by itself imply exploratory behavior, and it is here that the two frameworks separate. MOP has no parameter to tune and is adventurous by default; the comparison is sharpest at $d=0.8$, the precision at which the two agents have similar policy entropy (Figs.~\ref{fig:entropy_energy}a-c) and comparable lifespans. They nevertheless do not occupy the environment in the same way. MOP travels between the two food sources $13.3$ times per thousand steps against the Active Inference agent's $3.8$, and more often than Active Inference at any precision we tested; the same separation appears in where the time is spent, with the Active Inference agent standing most often on a food source or within one step of it, while MOP spreads its occupancy more evenly over the grid (Fig.~\ref{fig:exploration} in Appendix~\ref{app:temperature}). Fig.~\ref{fig:entropy_energy}e summarizes this as a single trade-off curve: MOP lies off the trace that Active Inference follows as $d$ is swept, since no precision matches its occupancy of the grid, and the one that comes closest in lifespan does so with a visibly narrower one. MOP therefore achieves exploration and survival jointly, whereas in Active Inference the two are exchanged against one another.

\section{Discussion}

We introduced a belief-based extension of the Maximum Occupancy Principle (MOP) and Active Inference to partially observable environments. The resulting framework generalizes naturally to POMDP settings via belief updating over hidden environmental variables. Our work introduces a novel reformulation of the Expected Free Energy (EFE) within the Sophisticated Inference framework. We express the EFE as a Bellman equation with a discount factor, enabling offline computation via value iteration over the full belief-state space, replacing the tree-search procedure of the original framework with a tractable dynamic programming solution that yields time-stationary policies directly applicable to POMDP settings. In our experiments, the induced policy of MOP exhibits structured adaptive behavior driven by internal energetic constraints. In particular, the agent balances exploratory behavior under high-energy conditions with increasingly goal-directed behavior under energy depletion, while maintaining flexible switching between available resources. The Active Inference agent, by contrast, exhibits robust survival behavior under uncertainty, driven by the joint minimization of risk and ambiguity that naturally balances food-seeking with information gathering about hidden food states. From a broader perspective, we identify a conceptual connection between MOP and Active Inference. Belief propagation under MOP is closely related to Variational Free Energy minimization for hidden-state inference.

Although our toy example is intentionally simple, we expect the qualitative similarities between the two frameworks to persist in more complex and stochastic environments. This connection has been discussed by Kiefer \cite{kiefer2025intrinsic}, who casts both frameworks as forms of constrained entropy maximization and argues that, in their full formulation, they differ mainly in whether the constraint that keeps the agent alive is explicit, as in Active Inference, or implicit, as in MOP. The ambiguity term in the EFE formulation (Eq. \ref{EFE_action}) drives the agent to seek observations that reduce uncertainty about hidden states. In the current grid-world, uncertainty is concentrated around food sources, leading the agent to revisit them to refine its beliefs. In richer environments with uncertainty distributed across many latent variables, an active inference agent would be naturally encouraged to explore broadly to reduce uncertainty. Therefore, even though active inference does not explicitly maximize state-action entropy in the same manner as MOP, it nevertheless induces exploratory behavior through epistemic value. The risk term in EFE plays a complementary role by enforcing homeostatic behavior: policies leading to preferred outcomes are favored, while trajectories associated with undesirable outcomes are avoided. At first glance, one may argue that such a survival-oriented mechanism is absent from MOP, since homeostatic constraints are not explicitly encoded as prior preferences. However, an analogous mechanism emerges implicitly through the treatment of terminal states. In practice, terminal states can be implemented as absorbing states that restrict the agent to a single self-transition action, eliminating all future behavioral possibilities. Since these states collapse the future trajectory distribution and reduce future path entropy to zero, they become strongly disfavored under MOP. Consequently, the agent is incentivized to avoid entering terminal states, producing behavior that resembles survival-driven action selection. From this perspective, both MOP and Active Inference can be interpreted as combining two complementary pressures: one favoring exploration and the maintenance of future possibilities, and another favoring continued viability or survival. Exploring this relationship further may help develop more unified and expressive models of adaptive behavior.


The two frameworks, however, arrive at these pressures by different means. Compared to fixed-horizon or preference-driven frameworks, MOP does not rely on engineered reward shaping or strongly specified priors, and no parameter tuning was performed to elicit the observed behaviors, hinting at the generalizability of these results. Moreover, Active Inference biases the agent toward confident (low-uncertainty) belief states through its prior preferences, whereas MOP does not encode such a bias, leading to a more neutrality-preserving belief evolution driven by path occupancy maximization. These results suggest that entropy-based control objectives provide a unifying perspective on intrinsic motivation in partially observable environments, bridging information-theoretic and free-energy based formulations.

\bibliographystyle{unsrt}
\bibliography{references}
\newpage
\appendix

\section{Appendix}
\subsection{Definition of various state spaces} \label{app:definitions}
We list the definitions of the terms used throughout: 
\begin{alignat*}{2}
x          &= (x_1, x_2)                 &\qquad& \textit{Spatial Position}\\
x_{f,i}          &= (1, 1) \; \text{or} \; (5,5)                 &\qquad& \textit{Food Source Position}\\
s          &= (x,E)                    &\qquad& \textit{Location and Energy (Observable) State}\\
h          &= (h_1,h_2)                  &\qquad& \textit{Hidden Food State}\\
(s,h)      &= (x,E,h_1,h_2)            &\qquad& \textit{World State}\\
\omega     &= (\omega_1,\omega_2)        &\qquad& \textit{Food Observation}\\
b(h)          &= (b(h_1),b(h_2))         &\qquad& \textit{Belief (about hidden food state)}\\
(s,b)      &= (x,E,b_1,b_2)            &\qquad&   \textit{Agent State (sufficient statistics)} \\
s'         &=  (x', E')&\qquad&  \textit{Next Observable State} \\
b'(h') &= (b'(h_1'), b'(h'_2)) &\qquad& \textit{Propagated Belief (one step forward)}
\end{alignat*}

\subsection{Environment Implementation}
\label{app:implementation}
\begin{table}[H]
\centering
\caption{Parameters used in the experiments.}
\label{tab:variables}
\begin{tabular}{ll}
\hline
Variable & Value \\
\hline
$E_{\max}$ & $30$ \\
$E_{\mathrm{gain}}$ & $15$ \\
Grid size & $5 \times 5$ \\
Food source locations & $(1,1)$ and $(5,5)$ \\
Initial $h_1$ & $1$ \\
Initial $h_2$ & $1$ \\
Initial $b(h_1=1)$ & $1$ \\
Initial $b(h_2=1)$ & $1$ \\
$\lambda$ & 0.8 \\
$\mu$ & 0.2 \\
$\rho$ & 0.8 \\
Belief discretization step & 0.1 \\
\hline
\end{tabular}
\end{table}
Results remain qualitatively unchanged for other values of $E_{\max}$, $E_{\mathrm{gain}}$ and the food-regeneration probabilities: these parameters modulate how conservatively the agents behave, but not the relative comparison between the three methods.




\subsection{Discretization of beliefs} \label{app:beliefs}
The belief state $b_i \in [0,1]$ is discretized onto a uniform grid with step size $\Delta$. The choice of $\Delta$ involves a trade-off: a smaller step yields a finer approximation but increases the number of belief states quadratically, since the agent maintains two independent beliefs $(b_1, b_2)$. At the same time, the total number of belief combinations scales as $(1/\Delta + 1)^2$, so reducing $\Delta$ from $0.1$ to $0.01$ increases the number of belief grid points by a factor of $100$. We therefore adopt $\Delta = 0.1$ as a practical choice that keeps the state space tractable while preserving the structure of the belief dynamics. Since the belief propagated by Eq.~\ref{belief} does not in general land on a grid point, we perform value iteration (Eq.~\ref{v_pomdp_opt}) on the grid and evaluate the successor value $V^*(s',b')$ by linear interpolation between the neighboring grid points at each iteration; for instance, for a propagated belief $b'_1 = 0.84$ the value $V^*(s',b')$ is obtained by interpolating linearly between the stored values $V^*(s', 0.8)$ and $V^*(s', 0.9)$.

\begin{figure}[t]
\centering
\includegraphics[width=0.7\linewidth]{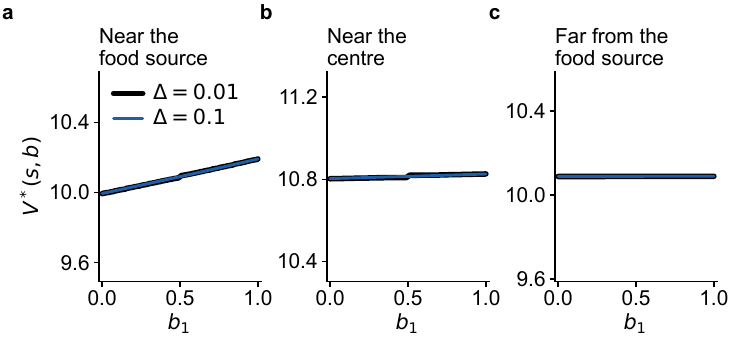}
\caption{Value function $V^*(s,b)$ as a function of belief $b_1$ (with $b_2=0.5$ fixed) for three representative states, computed with a fine grid ($\Delta=0.01$, black curve) and a coarse grid ($\Delta=0.1$, blue curve). The coarse grid points closely follow the fine curve in all cases, confirming that the $\Delta=0.1$ discretization introduces negligible approximation error.}
\label{fig:belief_discretisation}
\end{figure}

Figure~\ref{fig:belief_discretisation} shows the value function 
$V^*(s,b)$, computed via Eq.~\ref{v_pomdp_opt}, evaluated at three representative states as a function of $b_1$, with $b_2 = 0.5$ fixed, for both $\Delta = 0.1$ (coarse grid, black dots) and $\Delta = 0.01$ (fine grid, blue curve). In all three cases, the coarse grid points lie close to the fine curve, confirming that the discretization does not distort the value function — it merely quantizes it. The overall shape and magnitude of $V^*$ are preserved, and the differences between the two grids are negligible relative to the range of values. We therefore conclude that $\Delta = 0.1$ is a sufficient discretization for the purposes of this work.

\subsection{Beliefs in the two-food-source partially observable grid-world environment}
\label{app:beliefs_simple}
Using Eq. \ref{belief}, it is easy to check that when the agent is at the food source $i$, beliefs saturate
\begin{align*}
b'(h_i'=1|x'=x_{f,i},\;\omega_i=1)&=1\\
b'(h_i'=1|x'=x_{f,i},\;\omega_i=0)&=0 \;,
\end{align*}
because the observation $\omega_i$ is then fully informative about the hidden state $h_i$ of the food source $x_{f,i}$. Here, we have replaced the conditioning on $(s,b,a)$ with $x'=x_{f,i}$ explicitly, since the saturation of beliefs happens only when the next position $x'(s,a)$ coincides with food source $x_{f,i}$. In contrast, outside food source $x_{f,i}$, the belief diffuses to the prior as 
\begin{equation}\label{belief_nofood}
    b'(h_i') =\sum_{h_i} p(h_i'|x, h_i)\;b(h_i) \;.
\end{equation}

\subsection{Variational free energy and belief propagation} \label{VFE_belief}
The VFE is expressed as
\begin{equation*}
    F[Q] = D_{\mathrm{KL}}\big(Q(h')\,\|\,p(h')\big)
    - \mathbb{E}_{Q(h')}\big[\ln q(\omega' \mid s', h')\big],
\end{equation*}
where $Q(h')$ is the variational distribution over the updated hidden food state $h'$, $p(h') = \sum_h p(h'|h,x)\,Q(h)$ is the predictive prior induced by the transition dynamics, and $q(\omega' \mid s', h')$ is the observation model defined in Eq.~\ref{obs}. To prove that minimizing this equation is equivalent to belief propagation described in Sec.~\ref{section_beliefs}, we consider the two regimes induced by the structure of the observation model.

\paragraph{Case 1: Non-informative observations (agent away from food).}
When $x' \neq x_{f,i}$ for the corresponding food source, the 
observation model is uniform:
\[
q(\omega' \mid s', h') = \frac{1}{2}.
\]
Hence the accuracy term becomes independent of $h'$:
\[
\mathbb{E}_{Q(h')}[\ln q(\omega' \mid s', h')] = \ln \tfrac{1}{2},
\]
which is constant with respect to $Q(h')$. The VFE therefore reduces to
\[
F[Q] = D_{\mathrm{KL}}\big(Q(h')\,\|\,p(h')\big) + \text{const}.
\]
Minimization yields
\[
Q(h') = p(h') = \sum_h p(h'|h,x)\,Q(h),
\]
so that the posterior belief is entirely driven by the transition model. In this regime, belief updating reduces to pure prediction under the dynamics, recovering the propagation rule in Eq.~(\ref{belief_nofood}).

\paragraph{Case 2: Informative observations (agent at food location).}
When the agent is located at food source $x_{f,i}$ (i.e.\ $x' = x_{f,i}$), the observation model becomes deterministic as specified in Eq.~\ref{obs}, so observations perfectly identify the hidden state. In this case, the accuracy term becomes
\[
- \mathbb{E}_{Q(h')}[\ln q(\omega' |s', h')]
= - \sum_{h' \in \{0,1\}} Q(h')\,\ln q(\omega' | s', h').
\]
Since $q(\omega' \mid s', h')$ is degenerate, there exists a unique state $h^{\star}$ such that
\[
q(\omega' | s', h^{\star}) = 1, \qquad 
q(\omega' | s', h' \neq h^{\star}) = 0.
\]
If $Q(h' \neq h^{\star}) > 0$, the corresponding term in the expectation contains $\ln 0 = -\infty$, implying
\[
F[Q] = +\infty.
\]
Therefore, any finite minimizer must satisfy
\[
Q(h^{\star}) = 1, \qquad Q(h' \neq h^{\star}) = 0.
\]
Thus, the optimal posterior collapses to a point mass:
\[
Q(h') = \delta(h' - h^{\star}),
\]
which assigns full belief to the observed-consistent hidden state. In both cases, the resulting updates coincide with the belief propagation dynamics derived in Eq.~(\ref{belief_nofood}) and the deterministic update at food locations, respectively. The minimizer $Q^*(h')$ corresponds to the updated belief $b'(h')$ used throughout this work, i.e.\ $b'(h') = \arg\min_Q F[Q]$. Hence, variational free energy minimization provides a unified principle underlying the belief update equations in this grid-world POMDP.

\subsection{Empowerment} \label{app:mpow}
Consider a state $(s, b)$ at time $t$, which serves as the initial condition. The agent then executes an $n$-step action sequence $a^n = (a_t, \dots, a_{t+n-1})$, leading to a future observation state $o_{t+n} = (s_{t+n}, \omega_{t+n})$ at time $t+n$. The empowerment of state $(s,b)$ is defined as
\begin{equation} \label{mpow}
C(s,b)
=
\max_{p(a^n)}
\mathbb{E}_{P(a^n, o_{t+n} | s, b)}
\left[
\log
\frac{
P(o_{t+n} | a^n, s, b)
}{
P(o_{t+n} | s, b)
}
\right],
\end{equation}
where the expectation is taken over the joint distribution of action  sequences and future observations, conditioned on the initial state $(s,b)$ at time $t$. The marginal predictive distribution over future observations, obtained by averaging over all possible action sequences, is 
\begin{equation}
P(o_{t+n} | s, b)
=
\sum_{\tilde{a}^n}
P(\tilde{a}^n)\,
P(o_{t+n} | \tilde{a}^n, s, b).
\end{equation}
The predictive distribution over future observations factorizes as
\begin{equation}
P(o_{t+n} | a^n, s, b)
=
P(s_{t+n} | a^n, s, b)\;
P(\omega_{t+n} | s_{t+n}, b_{t+n}),
\end{equation}
where $s_{t+n} = (x_{t+n}, y_{t+n}, E_{t+n})$ is the future observable state reached after executing $a^n$ from $(s,b)$, $\omega_{t+n}$ denotes the observation generated at that state through the observation model (Eq.~\ref{obs}), and $b_{t+n}(a^n, s, b)$ is the belief propagated forward $n$ steps from the initial belief $b$ under action sequence $a^n$, updated at each intermediate step $k = t, \dots, t+n-1$ via the prediction step
\begin{equation}
b_{k+1}(h') = \sum_{h} p(h'|h, x_k)\; b_k(h),
\end{equation}
noting that intermediate observations are not conditioned upon, as empowerment considers all possible future trajectories rather than a single realized path. This factorization follows from the conditional independence of $\omega_{t+n}$ and $a^n$ given $s_{t+n}$ and $b_{t+n}$: once the future state $s_{t+n}$ and the propagated belief $b_{t+n}$ are known, the observation $\omega_{t+n}$ depends only on the hidden food state and the agent's location, and not on the specific action sequence that led there.

Computing the empowerment $C(s,b)$ requires maximizing the mutual information over all input distributions $p(a^n)$, which is a concave optimization problem solved here using the Blahut--Arimoto algorithm 
\cite{blahut1972computation} (see Appendix \ref{implementation}). 

Unlike the EFE and the value function $V^*$ of MOP, which both follow a Bellman structure and are computed via value iteration until convergence over the full state-belief space, empowerment does not admit a recursive decomposition of this form. Instead, for each state $(s,b)$ independently, all possible $n$-step action sequences are enumerated explicitly and their associated observation distributions are precomputed via forward belief propagation. The mutual information is then maximized over the input distribution $p(a^n)$ using the Blahut--Arimoto algorithm, which iterates until convergence for that specific state. The key distinction from value iteration is therefore that empowerment requires no information from neighboring states in the belief-state space — each state $(s,b)$ is solved as an entirely independent optimization problem. This makes the calculation exact for a fixed horizon $n$, but comes at a significant computational cost: the number of candidate action sequences grows as $|A|^n$, where $|A|$ is the number of available actions, so increasing $n$ by even one step multiplies the number of sequences by a factor of $|A|$. This exponential scaling imposes a practical upper limit on the horizon $n$, beyond which the computation of $C(s,b)$ becomes intractable.

\subsection{Implementation of MOP, EFE and Empowerment} \label{implementation}

\paragraph{MOP}
The value function for MOP is computed by solving the Eq. \ref{v_pomdp_opt} via value iteration, starting from an initial guess $V_0(s,b) = 0$ and iterating until the value function converges. The discount factor $\gamma$ controls the effective planning horizon: a value close to one weights distant future states almost as heavily as immediate ones, while a smaller value emphasizes short-term outcomes. We set $\gamma = 0.99$, which corresponds to an effective horizon of approximately $1/(1-\gamma) \approx 100$ steps into the future. We impose the boundary conditions $V(s^+, b) = 0$ for every terminal observable state $s^+$ (observable states $s$ with $E = 0$) regardless of $b$, so that terminal states have zero value due to the termination of the episode and the impossibility of generating any further action-state path entropy. The value iteration is terminated once the maximum change in the value function across all states falls below a threshold $\theta_{\mathrm{MOP}} = 10^{-2}$, which is usually reached after $N_{\mathrm{MOP}} = 250$ iterations.

\paragraph{EFE}
The EFE is computed via the same value-iteration procedure applied to the recursive expression in Eq.~\ref{EFE_action}, starting from $G_0(s,b) = 100$ and iterating until convergence. As in MOP, the discount factor $\gamma$ governs how far into the future the agent plans, and we use the same value $\gamma = 0.99$ for both schemes to ensure a fair comparison; this again corresponds to an effective horizon of roughly $100$ steps. The EFE additionally involves the inverse temperature $d$, which controls the stochasticity of the policy used in the recursive backup: as $d \to \infty$ the policy becomes deterministic (Eq.~\ref{optimal_EFE}), while finite $d$ yields a soft policy. The effect of $d$ is explained in detail in Appendix \ref{app:temperature}. The value iteration is terminated once the maximum change in $G(s,b)$ across all states falls below $\theta_{\mathrm{EFE}} = 10^{-2}$, reached after $N_{\mathrm{EFE}} = 300-400$ iterations, depending on $d$. 

\paragraph{Empowerment}
Computing $C(s,b)$ requires maximizing the mutual information over all input distributions $p(a^n)$, a concave optimization problem solved here using the Blahut--Arimoto algorithm \cite{blahut1972computation}. For a fixed state $(s,b)$, all $|A|^n$ action sequences are enumerated and the resulting observation distributions $P(o_{t+n}|a^n, s, b)$ are precomputed via forward belief propagation. The algorithm iteratively reweights the input distribution $p(a^n)$ to maximize the mutual information between action sequences and future observations, until convergence within a threshold $\theta_{\mathrm{BA}} = 10^{-5}$, typically reached after $N_{\mathrm{BA}} = 50$ iterations, yielding the optimal input distribution $p^*(a^n)$ and the empowerment value $C(s,b)$. This procedure is repeated independently for every state $(s,b)$ in the discretized belief-state space.

\subsection{Explaining the EFE equation}
\label{app:EFE_explain}
The per-action EFE is defined in Eq. \ref{EFE_action}, as
\begin{align*}
    G(s,b,a)
&=
\underbrace{
\mathbb{E}_{P(\omega', s', h'|s, b, a)}
\Big[
\overbrace{
\ln P(s'|s, b, a)
-\ln P(s')
}^{\text{Risk}}
\;
\overbrace{
-\ln q(\omega'|s', h')
}^{\text{Ambiguity}}
\Big]
}_{\text{Expected free energy of next action}}
\notag
\\
&\quad
+\gamma
\underbrace{
\mathbb{E}_{P(s', b'|s, b, a)}
\big[
G(s', b')
\big]
}_{\text{Expected free energy of subsequent actions}}
\end{align*}

Since the risk depends only on the marginal distribution over $s'$, it can be simplified by integrating $\omega'$ and $h'$ in the predictive model (Eq. \ref{predictive_hidden})
\begin{align*}
    \mathbb{E}_{P(\omega', s', h'|s, b, a)}\Big[
\ln P(s'|s, b, a)
-\ln P(s')\Big] 
&= \mathbb{E}_{P(s'|s, b, a)}\Big[
\ln P(s'|s, b, a)
-\ln P(s')\Big] \\
&= D_{KL} \Big[P(s'|s, b, a) \;\|\; P(s')\Big],
\end{align*}
where $D_{KL}$ denotes the KL divergence between the predicted distribution over future states and the agent's prior preferences. 
This term measures how far the agent expects to deviate from its preferred states, minimizing risk is equivalent to selecting actions 
that drive the agent towards states that conform to its prior preferences.

Similarly, the ambiguity term, for this environment, can be simplified
\begin{align*}
    \mathbb{E}_{P(s', \omega', h'|s, b, a)}\Big[
-\ln q(\omega'|s', h')\Big] \propto H \Big[q(\omega'|s', h')\Big],
\end{align*}
where $H$ denotes the Shannon entropy of the observation model. In the grid-world environment examined here, this term takes one of two values depending on the agent's location, following directly from the 
observation model in Eq.~\ref{obs}. When the agent is at a food source, $q$ collapses to a delta function and its entropy is zero — the observation is fully informative about the hidden food state. Away from a food source, the observation is completely uninformative and the entropy attains its maximum value, $H\bigl[q(\omega'|s', h')\bigr] = \ln \tfrac{1}{2}$. Minimizing ambiguity therefore drives the agent towards food locations, where it can resolve its uncertainty about the hidden state of the environment most effectively.

\subsection{Stochastic and deterministic policies of Active Inference} \label{app:temperature}
\begin{figure}[!t]
\centering
\includegraphics[width=0.7\linewidth]{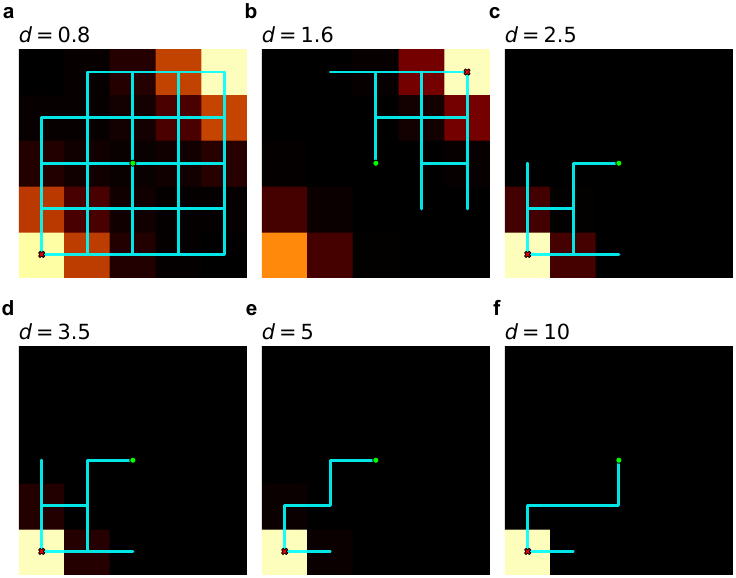}
\caption{Behavior of an active inference agent under stochastic policy, for various values of the inverse temperature, in the case of location-independent dynamics ($G^+=100$, 100 episodes per panel). One complete episode is drawn on each panel (start $\bullet$, end $\times$); the line retraces itself, so a committed agent appears as a few repeated edges rather than as a short path. In panels (b)--(f) the episodes are conditioned on the food source reached first, so that each heatmap shows a single commitment rather than a mixture of two.}
\label{temperature_fig}
\end{figure}

\begin{figure}[!t]
\centering
\includegraphics[width=0.7\linewidth]{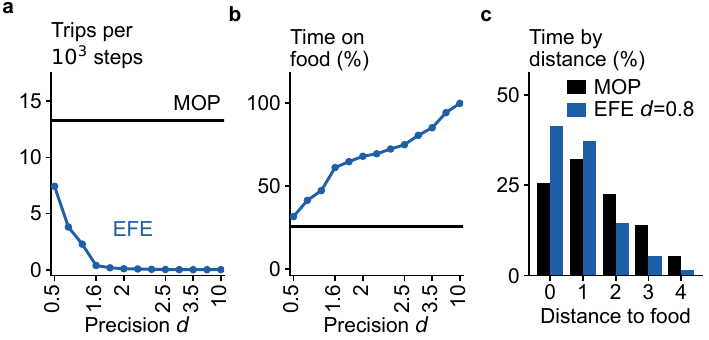}
\caption{\textbf{(a, b)} Per-episode behavior of the active inference agent across precision, 300 episodes per point; MOP (black) is drawn in both panels as a reference. \textbf{(a)} Number of trips between the two sources per $10^3$ steps, over the range of precision $d$. \textbf{(b)} Share of time spent standing on a food source. \textbf{(c)} Occupancy resolved by Manhattan distance to the nearest food source, for MOP and for the agent at $d=0.8$, the precision whose lifespan comes closest to MOP's: with lifespans matched, the active inference agent still spends $41\%$ of its time standing on a source against MOP's $26\%$, while MOP retains the occupancy at distances $3$ and $4$ that the active inference agent has largely given up.}
\label{fig:exploration}
\end{figure}

The inverse temperature parameter $d$, introduced in Eq.~\ref{G_policy}, controls the degree of stochasticity in the EFE policy and has a significant effect on the agent's behavior and survival. Fig.~\ref{temperature_fig} shows the spatial visitation heatmaps of the agent for six values of $d$, and Fig.~\ref{fig:exploration} quantifies the same behavior per episode, revealing three distinct regimes.

For $d \leq 1$ the policy is stochastic and the agent is explorative: it reaches a food source in every episode, visits both sources in $80$--$85\%$ of episodes, covers a median of $22$--$23$ of the $25$ grid cells, and travels between the two sources several times per thousand steps. Survival is shortest in this regime ($539 \pm 60$ steps at $d=0.5$). For $1 \leq d \leq 2$ the agent still uses both sources, but typically only once in a lifetime: at $d=1.6$ half of the episodes reach the second source, and they do so at a twentieth of the rate observed at $d=0.5$. For $d \gtrsim 2$ the agent commits to a single source and stays there: by $d=2.5$ only $5\%$ of episodes ever reach the other one, from $d=3$ almost none do, and occupancy of a single source rises from $68\%$ of the time at $d=2$ to $99.8\%$ at $d=10$, over a median of $6$ cells, with the longest survival of the range ($4883 \pm 381$ steps at $d=10$). While the normalized policy entropy is high (at $d=2.5$ the normalized policy entropy is still $0.60$ at $E=22$ (Fig.~\ref{fig:entropy_energy}b)), the direction in which the policy points becomes concentrated, and looks towards the food source.

MOP is drawn as a reference in each panel of Fig.~\ref{fig:exploration}. Of the three regimes, it most resembles the EFE agent at small $d$: it reaches both food sources in $94\%$ of episodes and spends only $26\%$ of its time standing on one (Fig.~\ref{fig:exploration}b), figures that only the lowest precisions approach. It nevertheless behaves differently. MOP moves between the two sources continually, $13.3$ trips per thousand steps, more than any EFE agent at any precision (Fig.~\ref{fig:exploration}a), and it spends far more of its time away from food. Fig.~\ref{fig:exploration}c makes this comparison at $d=0.8$, the precision with comparable policy entropy and lifespan: the EFE agent stands on a source $41\%$ of the time against MOP's $26\%$, while MOP spends $42\%$ of its time two or more steps from the nearest source against the EFE agent's $22\%$, nearly twice as much. Matched for how variable its policy is and for how long it lives, the EFE agent still hovers around food where MOP ranges away from it; food-source occupancy dominates its behavior across the whole range of $d$ tested.

\subsection{Boundary condition on the calculation of the EFE} \label{app:bound_cond}
The terminal boundary condition $G^+ = G(s^+, b)$, imposed on every terminal observable state $s^+$ (observable states with $E=0$) regardless of $b$, is the price the agent pays for dying, and it is not free to be chosen independently: the model already assigns a prior preference to $E=0$ through the target distribution $P(s)$ (Sec.~\ref{FEP}), and consistency with that preference fixes its value. Terminal states are absorbing, so the only path available from $s^+$ is the indefinite repetition of $s^+$ itself, and evaluating Eq.~\ref{EFE_action} along that path requires $G^+$ to satisfy its own Bellman equation,
\begin{equation} \label{G_plus}
G^+ = g^+ + \gamma \, G^+
\qquad \Longrightarrow \qquad
G^+ = \frac{g^+}{1-\gamma} = \frac{-\ln P(s\,|\,E{=}0)}{1-\gamma} \;,
\end{equation}
where $g^+ = -\ln P(s\,|\,E{=}0)$ is the risk incurred per step of occupying a terminal state, the self-transition being deterministic. With $P(s\,|\,E{=}0) = 10^{-6}$ and $\gamma = 0.99$ this gives $G^+ \approx 1382$. Read in the opposite direction, Eq.~\ref{G_plus} assigns to every choice of $G^+$ an implied preference for the terminal state,
\begin{equation} \label{G_plus_pref}
P(s\,|\,E{=}0) = \exp\big[-G^+ (1-\gamma)\big] \;,
\end{equation}
so that $G^+ = 0$ implies $P = 1$, death exactly as preferred as any other state; $G^+ = 100$ implies $P = 0.37$; and only $G^+ \approx 1382$ recovers the $10^{-6}$ that the risk term itself already uses.

\begin{figure}[H]
\centering
\includegraphics[width=0.7\linewidth]{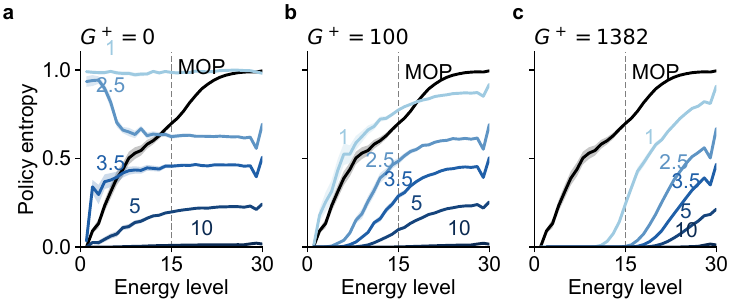}
\caption{Normalized policy entropy of the EFE agent against energy level, for three terminal boundary conditions $G^+$, with the MOP agent (black) as a fixed reference in each panel. MOP value function has no EFE terminal, so the same curve is the appropriate comparison throughout. Curves are labeled by the inverse temperature $d$.}
\label{fig:G_boundary}
\end{figure}
The choice is behavioral, not merely numerical (Fig.~\ref{fig:G_boundary}). With $G^+ = 0$ death carries no cost, and for every $d \lesssim 2.5$ the converged $G$ ranks a nearly dead state as better than a full one, so the agent is rewarded for starving; its policy entropy is flat in energy and close to uniform at low $d$ (panel a), giving behavior that is at once random and self-destructive. At the consistent value $G^+ = 1382$ the opposite holds: the agent becomes strongly risk-averse, its policy entropy collapses to near zero below $E \approx 10$--$15$ at every precision (panel c), and it survives far longer while occupying a much smaller part of the grid --- long-lived, but nearly deterministic and behaviorally impoverished. The intermediate value $G^+ = 100$ (panel b) is the most informative of the three: the ordering of $G$ in energy is correct, so the agent no longer seeks death, yet its policy remains genuinely stochastic and its entropy increases with energy as MOP's does, so that the agent both survives and explores. We therefore use $G^+ = 100$ throughout this paper. For large $d$ the three panels are nearly indistinguishable, since as $d \to \infty$ only the arg-min of $G(s,b,a)$ matters and the terminal enters as a nearly uniform offset across actions, leaving the deterministic agent insensitive to $G^+$. A larger $G^+$ also slows the convergence of value iteration, as the terminal value must propagate outward across the belief-state space before $G$ stabilizes.

\end{document}